\documentclass{article}
\usepackage[T1]{fontenc} 
\usepackage[utf8]{inputenc} 
\usepackage{amsmath,cite,url}
\usepackage{graphicx}
\usepackage[lbd]{ismir}
\usepackage[hidelinks,bookmarks=false]{hyperref}

\usepackage{color}

\usepackage[firstpage]{draftwatermark}
\definecolor{lightgray}{rgb}{0.9,0.9,0.9}
\definecolor{darkgray}{rgb}{0.4,0.4,0.4}
\SetWatermarkFontSize{12pt}
\SetWatermarkScale{1.1}
\SetWatermarkAngle{90}
\SetWatermarkHorCenter{202mm}
\SetWatermarkVerCenter{170mm}
\SetWatermarkColor{darkgray}
\SetWatermarkText{Late-Breaking / Demo Session Extended Abstract, ISMIR 2025 Conference}

\usepackage{lineno}

\title{Evaluating Fake Music Detection Performance Under Audio Augmentations}

\multauthor
{Tomasz Sroka \hspace{1cm} Tomasz Wężowicz \hspace{1cm} Dominik Sidorczuk} { \bfseries{Mateusz Modrzejewski \hspace{1cm}}\\
  Institute of Computer Science, Warsaw University of Technology, Poland\\
{\tt\small \{01169239, 01169258, 01169234, mateusz.modrzejewski\}@pw.edu.pl }
}

\def\authorname{D. Sidorczuk, T. Sroka, T. Wężowicz and M. Modrzejewski}

\begin{document}

\maketitle
\begin{abstract}

\begingroup
\setlength\parindent{0pt}
With the rapid advancement of generative audio models, distinguishing between human-composed and generated music is becoming increasingly challenging. As a response, models for detecting fake music have been proposed. In this work, we explore the robustness of such systems under audio augmentations.
\endgroup

To evaluate model generalization, we constructed a dataset consisting of both real and synthetic music generated using several systems. We then apply a range of audio transformations and analyze how they affect classification accuracy. We test the performance of a recent state-of-the-art musical deepfake detection model in the presence of audio augmentations. The performance of the model decreases significantly even with the introduction of light augmentations.

\end{abstract}
\section{Introduction}
Generative music models have made significant progress in recent years, allowing for the creation of synthetic music that closely resembles one composed by humans. In response, machine learning-based detectors are being designed to automatically recognize synthetic music and distinguish it from real compositions. Similar work has already been done in the field of detecting fake speech, for example \cite{xiao2025xlsrmambadualcolumnbidirectionalstate} and \cite{tak2022automaticspeakerverificationspoofing}. In this study, we investigate the performance of a recently proposed music model SONICS\cite{sonic:01}. The original paper demonstrates promising results on a prepared dataset consisting of real music and artificial music generated by Suno and Udio. Inspired by the insights from paper\cite{sonic:02} which highlights several important challenges in fake music detection - including robustness to audio modifications, ability to generalize across different generative models, model calibration, and interpretability, we concentrated our analysis on the first two aspects.

Our dataset contains music from Suno, Udio, YuE\cite{sonic:03} and MusicGen\cite{sonic:04} alongside real music samples. We then applied a variety of audio augmentations. The dataset and code used in this study are available on the Github repository\footnote{\url{https://github.com/Madghostek/wimu-sonics/}}. 

\section{Dataset}
The dataset consists of 20 songs from each model, including both the models used to train the SONICS model and new ones, to test its ability to generalize across different generative models. To generate the dataset, we used prompts generated with deepseek-r1 \cite{deepseekai2025deepseekr1incentivizingreasoningcapability} independently for every model. Then all audio files were initially downsampled to 16kHz as per model requirements.

\section{Experiments}
The original work offers a few different subtypes of the model. We have chosen ''SpecTTTra-$\alpha$'' with 120 seconds audio samples, which was the configuration that authors described as reaching the highest classification accuracy.
\subsection{Generalisation to different models}
Initially, we ran classification on our dataset as is, the results are shown in Table~\ref{tab:comparison}. The model doesn't generalize well, especially on MusicGen, which can be considered as sounding the most synthetic. Udio, despite being in the training dataset of the model, is still not classified correctly, as mentioned by the SONICS authors.

\begin{table}[h!]
 \begin{center}
 \begin{tabular}{|l|l|}
  \hline
  Model & Mean Probability of fakeness \\
  \hline
  Real  & $6.1\, (\pm4)\%$ \\
  \hline
  Suno  & $96.24\, (\pm5)\%$ \\
  \hline
  Udio  & $50.51\, (\pm43)\%$ \\
  \hline
  YuE  & $55.5\, (\pm32) \%$ \\
  \hline
  MusicGen  & $34.83\, (\pm31)\%$ \\
  \hline
 \end{tabular}
\end{center}
 \caption{Analysed model detections. Only real and Suno songs are classified correctly.}
 \label{tab:comparison}
\end{table}

\begin{figure*}
    \centering
    \includegraphics[width=1\linewidth]{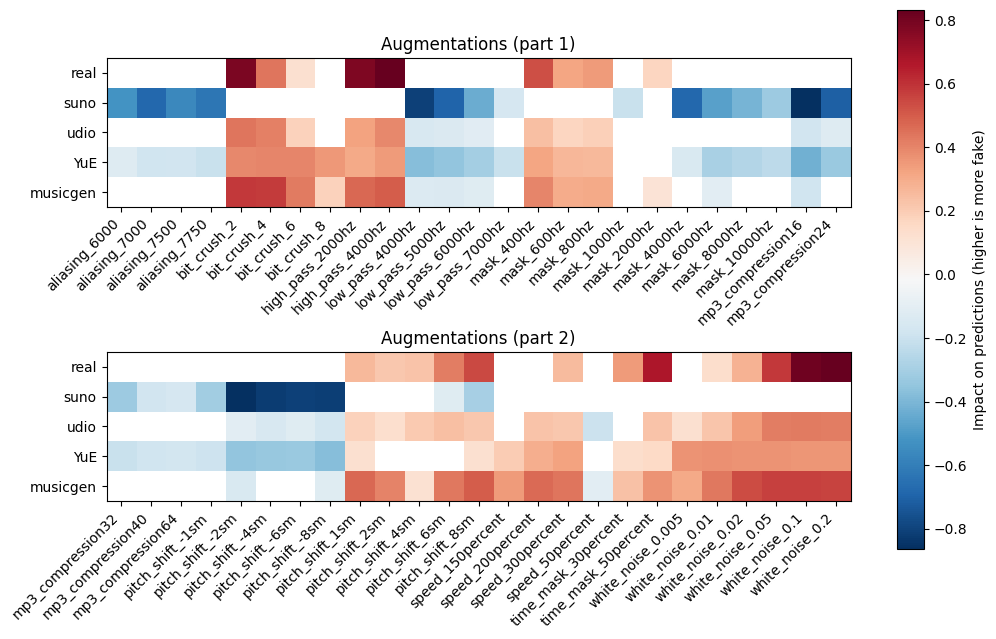}
    \caption{Various augmentations and their influence on predictions. Red entries mean the predictions are being pulled towards being fake.}
    \label{fig:heatmap}
\end{figure*}

\subsection{Robustness against augmentations}
We performed the following augmentations: aliasing, bit crush, equalization, high/low pass filtering, frequency masking, mp3 and ogg compression, pitch shifting, speed manipulation, silencing fragments of the waveform, reverb, vibrato, and white noise. The augmentations were tested across a range of parameters, but kept within a reasonable range not to make the audio incomprehensible. The specific configuration used to generate results for this paper can be found on the Github repository.

In Figure \ref{fig:heatmap} we only show augmentations with the most significant impact on model predictions. The colors describe difference between baseline predictions and new ones reported by the model (compare with Table~\ref{tab:comparison}). The most surprising is pitch shift, which depending on the shift direction can make any audio sound either highly genuine or fake. A shift down of two semitones already fools the model into classifying suno as highly real, while barely influencing the perception for a human. Another observation is that the more silence is being added to a song, the higher the probability of being fake is assigned by the model. In the extreme case, an empty file is classified as fake. This behavior might hinder classification even in a normal scenario. A solution might be to introduce a loss term during model training which ensures that silence is classified as ambiguous. We also observed a correlation between corrupting high frequencies of an audio and the probability of it being classified as fake. This might imply the model mistakenly learned to rely on certain artifacts present in the spectrum, instead of analyzing the audio as a whole. Some augmentations in the figure, especially low bit crush levels and strong white noise, skew the predictions are already highly perceptible for humans. The question whether the model should be able to handle such samples is hard to answer, because it's not obvious where to place the line of augmentations being too destructive.

\section{Conclusion and further work}

Firstly, we show that the SONICS model doesn't perform well on music from unseen generators. This is to be expected, since machine learning models are known for being unable to handle distribution shift between training and test data. Secondly, the results can be easily skewed by certain augmentations. The model almost always reacts to the modifications on music from different sources in consistent manner. This underlines the importance of utilizing diverse set of augmentations during training.

Our work can serve as a starting point for the analysis of different models, since the augmentation process is architecture agnostic, minus the sampling rate requirements. A further path to explore is the explainability of the predictions and understanding of the dependence on certain frequency bands by the model.

\bibliography{bib}

\end{document}